\def\final{1}

\documentclass[superscriptaddress, twocolumn, notitlepage, amsmath, amssymb, aps,
pra,
 10pt, letterpaper]{revtex4-1}

\usepackage{graphicx}% Include figure files
\usepackage{dcolumn}% Align table columns on decimal point
\usepackage{bm}% bold math

\usepackage{fancyhdr}
\usepackage{latexsym}
\usepackage{graphicx,color}
\usepackage{epstopdf}  %%after  \usepackage{graphicx,color}
\usepackage[pdfstartview=FitH]{hyperref}
\usepackage{xargs}
\usepackage{shadow,epsf,amsthm,amssymb,amsmath}

\usepackage{enumerate}

\usepackage{setspace}                           %for doublespacing

\usepackage{lipsum}% just to automatically generate text

\usepackage{tikz}	
\usetikzlibrary{backgrounds,fit,decorations.pathreplacing}  % TikZ libraries

\usepackage[colorinlistoftodos,prependcaption]{todonotes}
\newtheorem{definitionenv}{Definition}
\newtheorem{lemmaenv}[definitionenv]{Lemma}
\newtheorem{theoremenv}[definitionenv]{Theorem}
\newtheorem{corollaryenv}[definitionenv]{Corollary}
\newtheorem{propositionenv}[definitionenv]{Proposition}
\newtheorem{conjectureenv}[definitionenv]{Conjecture}
\newtheorem{remarkenv}[definitionenv]{Remark}
\newenvironment{remark}{\begin{remarkenv}\rm}{\end{remarkenv}}
\newcommand{\br}{\begin{remark}}
\newcommand{\er}{\end{remark}}

\newtheorem{exampleenv}{Example}
\newtheorem{app-lemmaenv}[section]{Lemma}

\newenvironment{definition}{\begin{definitionenv}\rm}{\end{definitionenv}}
\newenvironment{lemma}{\begin{lemmaenv}\rm}{\end{lemmaenv}}
\newenvironment{theorem}{\begin{theoremenv}\rm}{\end{theoremenv}}
\newenvironment{corollary}{\begin{corollaryenv}\rm}{\end{corollaryenv}}
\newenvironment{example}{\begin{exampleenv}\rm}{\end{exampleenv}}
\newenvironment{proposition}{\begin{propositionenv}\rm}{\end{propositionenv}}
\newenvironment{conjecture}{\begin{conjectureenv}\rm}{\end{conjectureenv}}
\newenvironment{app-lemma}{\begin{app-lemmaenv}\rm}{\end{app-lemmaenv}}

\newcommand{\bd}{\begin{definition}}
\newcommand{\ed}{\end{definition}}
\newcommand{\bl}{\begin{lemma}}
\newcommand{\el}{\end{lemma}}
\newcommand{\elp}{\hspace*{\fill} $\Box$
                 \end{lemma}}
\newcommand{\bt}{\begin{theorem}}
\newcommand{\et}{\end{theorem}}
\newcommand{\etp}{\hspace*{\fill} $\Box$
                 \end{theorem}}
\newcommand{\bc}{\begin{corollary}}
\newcommand{\ec}{\end{corollary}}
\newcommand{\ecp}{\hspace*{\fill} $\Box$
                 \end{corollary}}
\newcommand{\bcj}{\begin{conjecture}}
\newcommand{\ecj}{\end{conjecture}}

\newcommand{\be}{\begin{example}}
\newcommand{\ee}{\end{example}}
\newcommand{\eep}{\hspace*{\fill} $\Box$
                 \end{example}}
\newcommand{\bp}{\begin{proposition}}
\newcommand{\ep}{\end{proposition}}
\newcommand{\epp}{%\hspace*{\fill} $\Box$
                 \end{proposition}}

\newcommand{\bra}[1]{\langle#1|}
\newcommand{\ket}[1]{|#1\rangle}

\newcommand{\eeq}{ \setcounter{equation} {\value{enumi}}}

\newcommand{\ot}{\otimes}
\newcommand{\I}{\mathsf{id}}

\newcommand{\cC}{\mathcal{C}}
\newcommand{\cE}{\mathcal{E}}

\newcommand{\cG}{\mathcal{G}}

\newcommand{\cP}{\mathcal{P}}
\newcommand{\cQ}{\mathcal{Q}}

\newcommand{\cS}{\mathcal{S}}

\newcommand{\sfA}{\textsf{A}}
\newcommand{\sfB}{\textsf{B}}

\newcommand{\sfD}{\textsf{D}}

\newcommand{\sfF}{\textsf{F}}
\newcommand{\sfG}{\textsf{G}}
\newcommand{\sfH}{\textsf{H}}
\newcommand{\sfI}{\textsf{I}}
\newcommand{\sfM}{\textsf{M}}

\newcommand{\mZ}{{\mathbb Z}}

\renewcommand{\_}{\underline}

\def\beq{\begin{equation}}
\def\eeq{\end{equation}}

\def\bean{\begin{IEEEeqnarray*}{rCl}}
\def\eean{\end{IEEEeqnarray*}}

\ifnum\final=0
\newcommand{\mynote}[2]{{\color{#1} \marginpar{\tiny #2}}}
\newcommand{\mybignote}[2]{{\color{#1} $\langle \langle$ #2$\rangle \rangle$}}
\newcommandx{\rednote}[2][1=]{\todo[linecolor=red,backgroundcolor=red!25,bordercolor=red,#1]{#2}}
\newcommandx{\bluenote}[2][1=]{\todo[linecolor=blue,backgroundcolor=blue!25,bordercolor=blue,#1]{#2}}
\newcommandx{\yellownote}[2][1=]{\todo[linecolor=yellow,backgroundcolor=yellow!25,bordercolor=yellow,#1]{#2}}
\newcommandx{\greennote}[2][1=]{\todo[inline,linecolor=olive,backgroundcolor=green!25,bordercolor=olive,#1]{#2}}
\newcommand{\rmark}[1]{{\color{red} #1}}
\newcommand{\bmark}[1]{{\color{blue} #1}}

\else
\newcommand{\mynote}[2]{}
\newcommand{\mybignote}[2]{}
\newcommand{\rednote}[2][1=]{}
\newcommand{\bluenote}[2][1=]{}
\newcommand{\greennote}[2][1=]{}
\newcommand{\yellownote}[2][1=]{}
\newcommand{\rmark}[1]{#1}
\newcommand{\bmark}[1]{#1}

\fi

\begin{document}

\preprint{APS/123-QED}

\title{%Quantum Calderbank-Shor-Steane Stabilizer State Preparation by Classical Error-Correcting Codes
%Fault-tolerant  Preparation of Stabilizer States for Quantum CSS Codes
Fault-tolerant  Preparation of Stabilizer States for Quantum CSS Codes by Classical Error-Correcting Codes
}% Force line breaks with \\

\author{Ching-Yi Lai}
\email{cylai0616@iis.sinica.edu.tw}
\affiliation{\footnotesize Institute of Information Science, Academia Sinica, Taipei, Taiwan 11529}

\author{Yi-Cong Zheng}
%\email{zheng.yicong@quantumlah.org}
\affiliation{\footnotesize Centre for Quantum Technology, National University of Singapore, Singapore 117543}
\affiliation{\footnotesize Yale-NUS College, Singapore 138614}

\author{Todd A. Brun}%
%\email{tbrun@usc.edu}
 \affiliation{%
\footnotesize Electrical Engineering Department, University of Southern California, Los Angeles, California, USA  90089.\\
}%

\date{\today}% It is always \today, today,

\begin{abstract}
Stabilizer states are extensively studied in quantum information theory for their structures based on the Pauli group.  Calderbank-Shor-Steane (CSS) stabilizer states are of particular importance in their application to fault-tolerant quantum computation (FTQC).  However, how to fault-tolerantly prepare arbitrary CSS stabilizer states for general CSS stabilizer codes is still unknown, {and their preparation can be highly costly in computational resources. In this paper, we show how to prepare a large class of CSS stabilizer states useful for FTQC.  We propose distillation protocols using syndrome encoding by classical codes or quantum CSS codes.}  Along the same lines, we show  that classical coding techniques can {reduce the ancilla consumption in Steane syndrome extraction}  by using additional transversal controlled-NOT gates and classical computing power.  In the scenario of a fixed ancilla consumption rate, we can increase the frequency of quantum error correction and effectively lower the error rate.
\end{abstract}

\maketitle
\section{Introduction}%\todo[inline,size=\tiny,color=red!100!green!33]{You can use \textbf{bold in a comment}, like this.}
Quantum states are inherently susceptible to noise, and physical devices that process quantum information are themselves generally faulty. Reliable quantum computation is still possible, however, with the help of quantum error-correcting codes. {Quantum \emph{stabilizer} codes are an especially important class of quantum codes that are similar to classical linear block codes~\cite{Got97a}, in which} quantum information is encoded in the eigenstates---codewords---of a set of commuting Pauli operators called stabilizer generators.

{Fault-tolerant quantum computation (FTQC) is the task of accomplishing quantum computation with arbitrary accuracy using imperfect quantum circuits.  Protected by one or more stabilizer codes, a code-based FTQC scheme computes in the codespace of a stabilizer code, interspersed with repeated error corrections.}  {A \emph{fault-tolerant} procedure has the property that if only one component (or more generally, a small number of components) of the procedure fails, the errors produced by this failure are correctable, and are not transformed by the procedure into an uncorrectable error of the underlying error-correcting code.}  Threshold theorems have shown that it is possible to realize quantum computations of arbitrary size with arbitrary accuracy, provided that the errors are sufficiently local and their rates fall below a threshold~\cite{AB97,DS96,Gaitan08,LB13}.  Currently, most FTQC schemes use Calderbank-Shor-Steane (CSS) type stabilizer codes~\cite{CS96,Ste96a},
where every stabilizer generator can be chosen to be the  tensor products of identity and either  $X$ or $Z$ Pauli operators, and so can the logical operators.  {Most such FTQC schemes require the preparation of CSS stabilizer states---codewords of the CSS code that are eigenstates of some set of logical operators---for the purpose of error correction or computation.}

CSS stabilizer states can be  prepared by using Clifford encoding circuits (with faulty gates)~\cite{NC00}.  However, this is not fault-tolerant, so the generated states need to be verified.  Basic CSS stabilizer states, such as the logical states $\ket{0}_L$ or $\ket{+}_L$, are usually fault-tolerantly generated by specific quantum circuits  in FTQC schemes~\cite{AGP06:QAT:2011665.2011666,Knill05nature,AC06:PhysRevLett.98.220502,PR12,SDT07,SR08,LPSB13,FSG09,Ste02,Goto16}.  For general CSS codes, it is  not known how to produce arbitrary stabilizer states that are ``clean'' enough for quantum computation, especially when the code length is large.

In other contexts (e.g., entanglement purification \cite{BDSW96,BBPSSW96} or {magic state distillation} \cite{BK05}), this problem is tackled by distillation:  making a bunch of imperfect states, and then carrying out a protocol to produce a smaller number of better states.  {In this paper we show how \emph{classical error-correcting codes}, together with the Steane syndrome extraction, can be applied to distill a large class of useful CSS stabilizer states (Distillation Protocol~I), by actively correcting errors on a fraction of the imperfect stabilizer states that are produced by non-fault-tolerant (or fault-tolerant) methods.

If we have clean ancillas, we can use the Steane syndrome extraction to learn information about the errors---the \emph{error syndrome}~\cite{Ste97L}.  A transversal circuit is applied between the codeword and two clean ancillas, and bitwise qubit measurements are applied to the ancillas and the error syndrome is obtained by computing the parities of the corresponding measurement outcomes.  However, this would obviously consume more clean ancillas than it produces.  Since in our scenario only noisy stabilizer states are available, we combine  Steane extraction with classical coding.  After performing a transversal circuit on a set of noisy CSS stabilizer states, a subset of the states are measured bitwise and a set of parities is calculated, and classical decoding is then applied to this set of parities to learn the error syndromes of the remaining stabilizer states.  Quantum error correction can be applied accordingly to obtain clean ancillas.}  Along the way, we also develop a distillation protocol using quantum CSS codes rather than classical codes (Distillation Protocol~II).

The only operations needed in the two protocols are \emph{transversal} controlled-NOT (CNOT) gates, bitwise single-qubit measurements, classical decoding, and correction by applying Pauli gates (see Sec.~\ref{sec:distillation}).  These features for fault-tolerance are similar to the constraint of local operations and classical communication (LOCC) in some multipartite  protocols, {where each qubit is considered as a single party}.  Therefore our distillation protocols naturally apply to the task of multipartite entanglement purification for CSS stabilizer states by LOCC~\cite{DAB03,ADB05,CL04,HDD06,KMBD06,GKV06}.  {For simplicity, we only consider CSS codes that encode one logical qubit in this paper. Our results, however, can be generalized easily to multi-qubit codes.}

{The methods used for distilling ancillas---combining Steane syndrome extraction with classical error correction---can also solve a different (but related) problem.}  Steane syndrome extraction is used for quantum  CSS codes with high-weight stabilizer generators.  \rmark{However, each error-correction step requires two clean ancillas per (quantum) codeword, which are of the same size as the underlying CSS codes and are expensive,  especially when the code length is large.}  {We therefore would like to use as few ancillas as possible during syndrome measurement without seriously degrading the performance of error correction.  To achieve this, we propose an \emph{ancilla saving} protocol using classical codes.  Rather than using two ancillas for each code block, a smaller number of ancillas is shared among multiple code blocks, and classical decoding is used to separate out the error syndromes of the different blocks.  Assuming that the error rate is low enough, this can reduce the rate of ancilla consumption for a given rate of error correction without seriously reducing its accuracy.}

The paper is organized as follows. We provide preliminary material in the Sec.~II, including the basics of stabilizer codes, CSS codes, and Steane syndrome extraction.  The distillation protocols by classical codes and quantum CSS codes are given in Subsec.~\ref{sec:classical distillation} and Subsec.~\ref{sec:quantum distillation}, respectively.  Following that, we describe the ancilla saving protocol in Sec.~\ref{sec:ancilla saving}.  We conclude in Sec.~\ref{sec:discussion}.

\section{Preliminaries}

We begin with a brief review of classical codes, quantum stabilizer codes, CSS codes, and  Steane syndrome extraction.

\subsection{Classical Codes, Stabilizer Codes, and CSS Codes}

{Error-correcting codes protect digital information from noise by adding redundancy.}  The encoded information has to satisfy some mathematical relations---{\emph{parity checks}}---so that errors can be detected if any of the parity checks are violated.  {Let $\sfH$ be an $(m-k)\times m$ binary matrix with full rank.} Then an $[m,k,d]$ linear binary code $\cC$ associated with parity-check matrix $\sfH$ is a $k-$dimensional subspace of all binary ordered $m-$tuples (row vectors) in $\mathbb{Z}_2^m$ such that
\[
v\sfH^T=0,
\]
for all $v\in\cC$, where $H^T$ is the transpose of $H$ and the addition is modulo $2$.  Such row vectors $v$ are called \emph{codewords} of $\cC$.  If $\sfH \tilde{v}^T\neq 0$ for some $\tilde{v}\in \mathbb{Z}_2^m$, we know that some error occurred.  Hence, the rows of $\sfH$ are called parity-checks and $\sfH \tilde{v}^T$ is called the\emph{ error syndrome} of $\tilde{v}$.  The parameter $d$ is called the minimum distance of $\cC$ so that any two codewords of $\cC$ differ in at least $d$ bits.  This code can correct arbitrary $\lfloor \frac{d-1}{2}\rfloor$-bit errors.  Since the code is linear, $\cC$  can also be defined as the row space of an $k\times m$  generator matrix $\sfG$, which satisfies
\[
\sfG\sfH^T=0.
\]
{That is, $\sfG$ and $\sfH$ are orthogonal.}  The dual code $\cC^{\perp}$ of $\cC$  is the row space of $\sfH$.  For more properties of classical codes, please refer to \cite{MS77}.

Now we consider the quantum case. {We focus on the two-level quantum system---the \emph{qubit}. A pure qubit state is a unit vector in the two-dimensional complex vector space $\mathbb{C}^2$ with the usual inner product and an (ordered) orthonormal basis $\{\ket{0},\ket{1}\}$.}  The $n$-qubit state space is  $\mathbb{C}^{2^n}$.  Let $\mathcal{G}_n = \mathcal{G}_1^{\otimes n}$
denote the $n$-fold Pauli group, where
\begin{equation}
\mathcal{G}_1=\{\pm I, \pm iI, \pm X, \pm iX, \pm Y, \pm iY, \pm Z, \pm iZ\} ,
\end{equation}
and
\[
I=\begin{bmatrix}1 &0\\0&1\end{bmatrix},\ \ \  X=\begin{bmatrix}0 &1\\1&0\end{bmatrix},
\]
\[
Z=\begin{bmatrix}1 &0\\0&-1\end{bmatrix},\ \ \   Y = iXZ .
\]
We use the notation $X_j$ to denote $I^{\otimes j-1}\otimes X\otimes I^{\otimes n-j}$ where the underlying length $n$ is clear from the context, and similarly for $Y_j$ and $Z_j$. We also use the notation $X_e$ for $e=e_1\cdots e_n\in \mathbb{Z}_2^n$ (row vector) to denote $\otimes_{i=1}^n X^{e_i}$, and similarly for $Z_e$.
\rmark{In general, an $n$-fold Pauli operator can be expressed as
\begin{align}
i^{c'} \otimes_{i=1}^n X^{e_i}Z^{f_i}= i^c X_e Z_f\label{eq:pauli}
\end{align} for $c,c'\in\{0,1,2,3\}$ and $e,f\in \mathbb{Z}_2^n$.
Thus $(e,f)$ is called the binary representation of the Pauli operator $i^c X_e Z_f$.}
For example, $I\ot X\ot I\ot I\ot Z= X_2Z_5= X_{01000}Z_{00001}$.

The identity $I^{\ot n}$ will be denoted by $\I$.  Let $C_i(X_j)$  denote the CNOT gate with control qubit~$i$ and target qubit~$j$.  For example, $C_1(X_2)= \ket{0}\bra{0} \ot I^{\ot n-1}+ \ket{1}\bra{1}\ot X\ot I^{\ot n-2}$.  The quantum circuits in this paper consist only of CNOT gates, together with single-qubit measurements in the $X$ or $Z$ basis.

Suppose  $\mathcal{S}$ is an Abelian subgroup of   $\mathcal{G}_n$ with a set of $l$ independent generators  $\{g_1,\dots, g_{l}\}$, and $\mathcal{S}$ does not include $-I$.  Every element in $\cG_n$ has eigenvalues $\pm 1$.  An $[[n,n-l]]$ quantum stabilizer code $C(\mathcal{S})$ is defined as the $2^{n-l}$-dimensional subspace of the $n$-qubit state space ($\mathbb{C}^{2^n}$) fixed by  $\mathcal{S}$, which is the joint-$(+1)$ eigenspace of $g_1, \dots, g_{l}$.  Then for a codeword $\ket{\psi}\in C(\cS)$,
\[
g\ket{\psi}=\ket{\psi}
\]
for all $g\in \mathcal{S}$.  {When $l=n$, $C(\cS)$ has only one eigenstate (up to a phase) and this state is called a \emph{stabilizer state}.}

{The error operators in this paper are assumed to be \emph{Pauli errors} (i.e., operators in $\cG_n$).  This is not actually as restrictive as it sounds:  since the Pauli operators form a basis, the ability to correct a set of Pauli errors implies the ability to correct a large class of general errors.}  If a Pauli error occurs on $\ket{\psi}$, some eigenvalues of $g_1,\dots, g_{l}$ may be flipped.
Therefore, we gain information about the error by measuring the stabilizer generators $g_1,\dots, g_{l}$,
and the measurement outcomes (in bits) of $g_1,\dots, g_{l}$ are called \emph{error syndrome}.  (If the eigenvalue of a stabilizer is $+1$ or $-1$, its corresponding syndrome bit is $0$ or $1$, respectively.)  Then a quantum decoder has to choose a good recovery operation based on the measured error syndromes.

{CSS codes are a class of stabilizer codes whose stabilizer generators consist of the tensor products of identity and either $X$ or $Z$ operators~\cite{CS96,Ste96a}.}  Let $[\sfM]_{i,j}$ denote the $(i,j)$ entry of a matrix $\sfM$.  (We may also use $[v]_i$ to denote the $i$th entry of a vector $v$.)  A CSS code can be defined by two matrices that are orthogonal to each other.  Suppose $\sfH_Z$ and $\sfH_X$ are $r_Z\times n$ and $r_X\times n$ matrices $(r_Z+r_X\leq n)$ with full rank $r_Z$ and $r_X$, respectively, such that
\begin{equation}
\sfH_X \sfH_Z^T =0.
\label{eq:matrixOrthog}
\end{equation}
Then we can define an $[[n,n-r_Z-r_X]]$ CSS code with $Z$ stabilizer generators
\begin{equation}
g_i =\bigotimes_{j=1}^n Z^{[\sfH_Z]_{i,j}}, \quad i=1,\dots, r_Z,
\end{equation}
and $X$ stabilizer generators
\begin{equation}
g_{r_Z+i} =\bigotimes_{j=1}^n X^{[\sfH_X]_{i,j}}, \quad i=1,\dots, r_X.
\end{equation}
The condition in Eq.~(\ref{eq:matrixOrthog}) implies that the $X$ and $Z$ stabilizer generators all commute.

\rmark{The \emph{check matrix} of an $[[n,n-l]]$ stabilizer code is an {$l \times 2n$} binary matrix whose rows are the binary representations (Eq.~(\ref{eq:pauli})) of the stabilizer generators $g_1,\dots, g_{l}$.}
For the  CSS code defined by $\sfH_Z$ and $\sfH_X$, its check matrix is
\begin{equation}
\begin{bmatrix}\sfH_Z&0\\
0&\sfH_X\end{bmatrix}.
\end{equation}

 \rmark{The error syndrome of a Pauli error is the string of binary outcomes of measuring the stabilizers $g_1,\dots, g_{l}$.
Since a Pauli error can be expressed as a product of $X$ and $Z$ operators,
error correction can be done by  treating Pauli $X$  and $Z$ errors separately.
%, but similarly, when decoding a CSS code.
For CSS codes, the eigenvalues of $g_1,\dots, g_{r_Z}$ (respectively, $g_{r_Z+1},\dots, g_{l}$) correspond to the error syndrome of $X$ (resp. $Z$) errors.
Suppose a Pauli error $X_{e}Z_f$ occurs on a codeword, where $e,f\in \mathbb{Z}_2^n$.}
 %are binary $n$-tuples indicating which qubits have $X$ and $Z$ errors, respectively.
Then its (binary) $X$ error syndrome $s_X\in\mathbb{Z}_2^{r_Z}$, which corresponds to the eigenvalues  of $g_1,\dots,g_{r_Z}$, is given by
\begin{equation}
s_{X}=e\sfH_Z^{T} ,
\end{equation}
and the $Z$ error syndrome is defined similarly:
\begin{equation}
s_Z= e\sfH_Z^T .
\end{equation}

{In addition to the stabilizer elements, there will in general also be Pauli operators that commute with all the stabilizer generators without being in the stabilizer themselves.  We call these {\it logical operators}.  For a CSS code, it is always possible to find a subset of logical operators involving only the identity and either $X$ or $Z$ operators, which generate all other logical operators up to multiplication by a stabilizer element.  Moreover, these logical generators can be chosen in anticommuting pairs, usually denoted $\bar{X}_j$, $\bar{Z}_j$, where each pair corresponds to logical qubit $j$ in the code.}
{For simplicity,  in the rest of this paper we consider the preparation of stabilizer states of CSS codes $\cQ$ that encode one logical qubit (or $r_Z+r_X=n-1$).  Our results can be directly generalized to the case of $[[n,n-r_Z-r_X]]$ multiple-qubit  CSS codes $(n-r_Z-r_X>1)$.  Let $\bar{X}$ and  $\bar{Z}$ denote the logical $X$ and $Z$ operators of $\cQ$.  A quantum state with $L$ in the subscript refers to an encoded state. For example, the encoded $\ket{0}$, $\ket{+}=\frac{1}{\sqrt{2}}(\ket{0}+\ket{1})$ are denoted by $\ket{0}_L$ and $\ket{+}_L$, respectively.}

\subsection{CSS Stabilizer States and Steane Syndrome Extraction}

\begin{figure}
\includegraphics[scale=0.3]{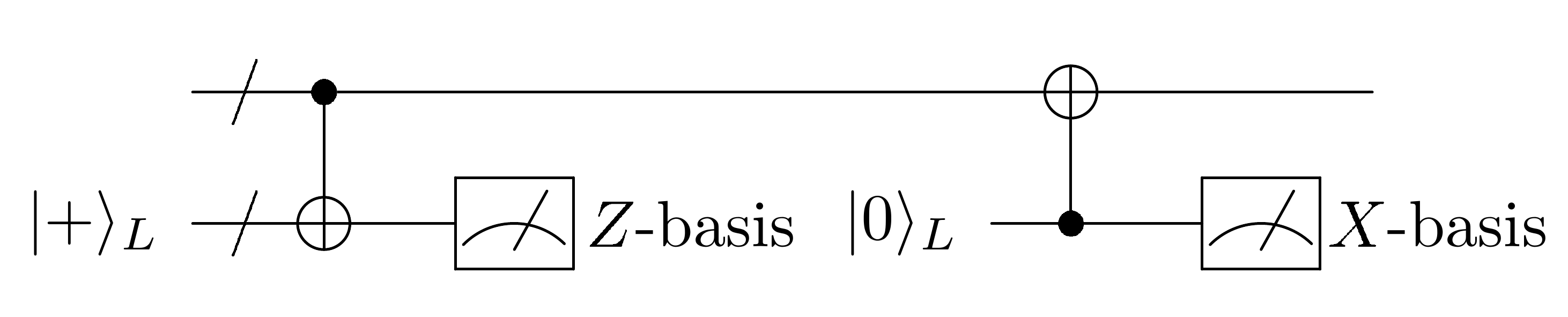}
\vspace{-0.2cm}
\caption{Quantum circuit for Steane syndrome extraction.}
\label{fig:steane syndrome}
\end{figure}

Steane suggested a method to extract error syndromes for CSS codes~\cite{Ste97L}, as shown in Fig.~\ref{fig:steane syndrome}.  Two clean ancillas $\ket{+}_L$ and $\ket{0}_L$ in the logical states of the underlying CSS code $\cQ$ are used to measure the $X$  and $Z$ error syndromes, respectively.  Each  CNOT gate  in Fig.~\ref{fig:steane syndrome} represents transversal CNOT gates, and the measurements in the $Z$ or $X$ basis are bitwise.  In the circuit, $X$ and $Z$ errors on the data qubit will propagate, respectively, to the ancillas $\ket{+}_L$ and $\ket{0}_L$  through the CNOTs, so that we learn error information by measuring the two ancillas.

Suppose the measurement outcomes of $\ket{+}_L$ and $\ket{0}_L$ are $m_X$ and $m_Z$ (in bits), respectively.
\rmark{Then the (observed) $X$ and $Z$ syndromes are computed by
$
m_X\sfH_Z^{T},$   and  $m_Z \sfH_X^{T},
$
respectively.}  We can perform error correction according to these syndromes or just keep track of them.

The two ancillas $\ket{+}_L$ and $\ket{0}_L$ are actually stabilizer states of $\cQ$ by including logical operator $\bar{X}$ or $\bar{Z}$ in with the stabilizer generators.  That is, $\ket{+}_L$ is stabilized by $\langle g_1, \dots, g_{n-1}, \bar{X}\rangle$, and $\ket{0}_L$ is stabilized by $\langle g_1, \dots, g_{n-1}, \bar{Z}\rangle$.  {These two stabilizer states can be produced by Clifford encoding circuits with CNOT gates only, together with the ability to prepare physical qubits in $\ket{0}$  and $\ket{+}$~\cite{Gra02}.}  If imperfect quantum gates are used in the encoding circuit, the states must be verified.

We can  use  Steane syndrome extraction to correct errors on a noisy ancilla, but it requires us already to have two clean ancillas, which is clearly impractical.  In the following section, we will introduce a protocol for distilling a number of CSS stabilizer states from a larger set of imperfect ones by using classical error-correcting codes and quantum CSS codes, and demonstrate how to distill the states $\ket{0}_L$ or $\ket{+}_L$.

A simple example of a CSS stabilizer state is the  Einstein-Podolsky-Rosen (EPR) pair:
\[
\frac{\ket{00}+\ket{11}}{\sqrt{2}},
\]
which is stabilized by $X\otimes X$ and $Z\otimes Z$.  {C{\'o}rcoles} \emph {et~al.} recently experimentally demonstrated error correction on an EPR pair~\cite{CMSCSGC15}.  EPR pairs were the first states for which a distillation protocol was proposed \cite{BDSW96,BBPSSW96}.

\section{Ancilla distillation}
\label{sec:distillation}

Suppose we are using an $[[n,1]]$ CSS code $\cQ$ defined by matrices $\sfH_Z,\sfH_X$ with $r_Z+r_X= n-1$ as in the previous section.  Given a noisy stabilizer state, e.g., $X_e\ket{{0}}_L$, we can perform quantum error correction and restore the state up to some logical operator if we know the actual error syndrome.  Now, measuring the logical operator $\bar{Z}$ will tell us whether there is a logical error or not, and additional logical correction can be applied if necessary.  Thus we can, ideally, have a \emph{perfect} stabilizer state $\ket{{0}}_L$.  In reality we are in a situation where the ancillas for syndrome measurements are also imperfect, and we will address this issue in this section.

Suppose we are given a bunch of imperfect ancillas in some CSS stabilizer state, e.g., $\ket{0}_L$ or $\ket{+}_L$, and we wish to purify them.  Our approach is to determine the correct error syndromes of a  subset of the ancillas by measuring the rest.  More precisely, we will use a transversal quantum circuit to couple $m$ noisy ancillas according to a classical error-correcting code, measure $m-k$ of them, and then extract the error syndromes of the remaining $k$ ancillas. This procedure is called \emph{ancilla distillation by classical codes} and  will be detailed in the first subsection.  In the second subsection, we generalize the idea to distill the noisy ancillas by using an arbitrary quantum CSS code.

For now, we neglect any errors in the CNOTs or measurements used in the quantum circuit for distillation.  We discuss the issue of noisy distillation circuits in Section \ref{sec:discussion}.

\subsection{Distillation by Classical   Codes}
\label{sec:classical distillation}

The key observation is that classical binary linear codes can be encoded or decoded by circuits using only CNOTs.  Suppose $\cC_D$ is an $[m,k,d]$ binary linear block code  that can correct $t=\lfloor \frac{d-1}{2}\rfloor$ errors.  Such a code has $r = m-k$ parity checks.  Let $\sfH_D= [  \sfA^T \ \sfI_r ]$ be the parity-check matrix of $\cC_D$ in systematic form, where  $\sfI_k$ is the $k\times k$ identity matrix and $\sfA$ is $k\times r$.  We define a quantum distillation circuit $U_D$ by
\begin{align}
U_D=\prod_{i=1}^k\prod_{j=1}^r C_i(X_{k+j})^{[\sfA]_{i,j}}.
\end{align}
That is, $C_i(X_{k+j})$ is applied if $[\sfA]_{i,j}=1$, and $\I$ is applied, otherwise.  Consider $X_e \ket{0^m}=\ket{e}$, where $e=e_1\cdots e_m\in \mathbb{Z}_2^m$ and
\[
0^m=\underbrace{0\cdots 0}_m .
\]
Then
\[
U_D \ket{e}= \ket{e_1\cdots e_k}\ot\ket{s_e},
\]
where $s_e^T= \sfH_D e^T$ is the classical error syndrome of $e$ with respect to $\cC_D$.  Then we can use a decoder of $\cC_D$ to find the most probable error vector $\tilde{e}\in\mZ_2^m$ and then correct the bit-flip errors in $\ket{e_1\cdots e_k}$.  This decoding procedure is the main conceptual tool of our distillation protocol.

Distillation of CSS stabilizer states by classical codes involves two rounds of  error correction: one for $X$ errors and one for $Z$ errors.
\medskip

\noindent\emph{Distillation Protocol I:}
\begin{enumerate}[1)]
\item  Using an encoding circuit, we prepare many noisy copies of an $n$-qubit CSS stabilizer state.  Divide the noisy ancillas up into groups of~$m$.

\item \label{step:I2} {(Round~1: $X$ errors)} %\cnote{a better name here?}
In each group of $m$ ancillas,  choose the last $r$ of the ancillas to hold the parity checks, and apply $U_D$ {\it transversally}:  that is, apply $U_D$ to the first qubits of all $m$ ancillas in the group, to the second qubits, and so forth.  This unitary $U_D$ applies transversal CNOTs according to the pattern of 1s in the binary matrix $A$.

\item Measure every qubit in the $r$ parity-check ancillas (the last $r$ ancillas in each group of $m$) in the $Z$ basis. {Let the binary row vectors $\nu^{\tiny (1)},\dots, \nu^{\tiny (r)}\in \mZ_2^n$ be the outcomes of these measurements.}

\item Calculate $\sigma^{\tiny (i)}\triangleq \nu^{\tiny (i)}  \sfH_Z^T$ for $i=1,\dots,r$.  For $j=1, \dots, r_Z$, use $[\sigma^{\tiny (1)}]_j \cdots [\sigma^{\tiny (r)}]_j$ as a classical error syndrome of $\cC_D$ and use a decoder of $\cC_D$ to find the most probable error vector $\tilde{s}^{\tiny (1)}_j \cdots \tilde{s}^{\tiny (m)}_j$ with this error syndrome.  Then $\tilde{s}^{\tiny (i)}_1\cdots \tilde{s}^{\tiny (i)}_{r_Z}$ is the estimated $X$ error syndrome of the $i$th target ancilla.  Correct the $X$ errors, if any (or just keep track of them).

\item If the distillation target is $\ket{0}_L$ or $\ket{1}_L$, calculate the parity of $\bar{Z}$ from $\nu^{\tiny (i)}$ and estimate the syndrome bits for the logical operators as in the previous step.
Correct the logical errors $\bar{X}$, if any (or just keep track).

\item Of our original large number of ancillas, a fraction $k/m$ are left.  Again, divide them up into groups of $m$. It is very important that ancillas that were grouped together in the first round are not grouped together in the second round, because their errors are correlated.

\item {(Round~2: $Z$ errors)} Similarly to step 2 above, do a transversal $U_D^H$, where
\begin{align}
U_D^H=\prod_{i=1}^k\prod_{j=1}^r C_{k+j}(X_{i})^{[\sfA]_{i,j}}.
\end{align}
(The control and target qubits of $U_D$ are switched to obtain $U_D^H$.)

\item Measure the $r$  parity-check ancillas in the $X$ basis. Then repeat steps 3--5, but with $X$ and $Z$ switched everywhere, and with $\ket{0},\ket{1}$ replaced by $\ket{+},\ket{-}$, respectively.
\hspace*{\fill} $\Box$
\end{enumerate}

\begin{figure}
\includegraphics[scale=0.20]{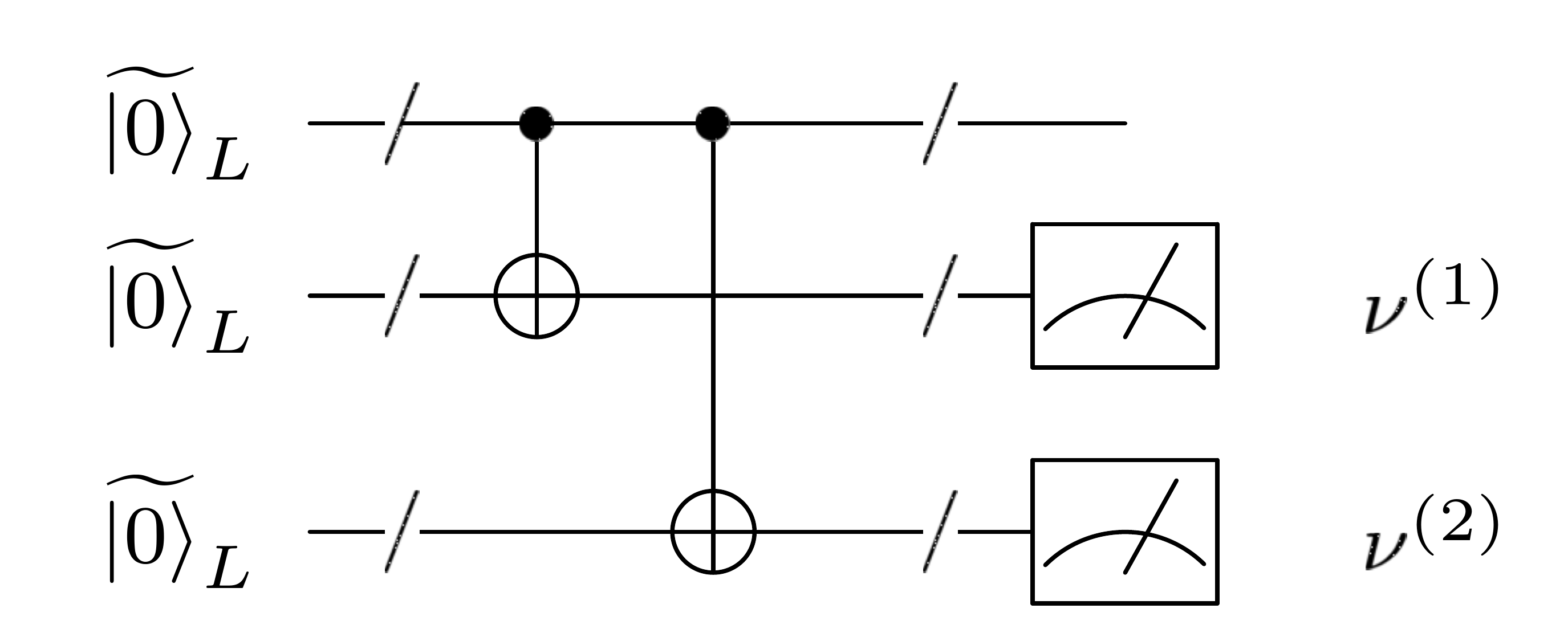}
\caption{The circuit for distilling $\ket{0}_L$   by the classical $[3,1,3]$
repetition code, where  $\widetilde{\ket{0}}_L$ are imperfect logical states. The last two $\widetilde{\ket{0}}_L$ serve as parity-check ancillas.}\label{fig:distillation_repetition code}
\end{figure}

%\rmark{This procedure is somewhat technical, so we will  demonstrate this distillation protocol with an example in detail.}
This procedure is somewhat technical, so we will demonstrate the distillation protocol with a detailed example.

 \be
 Fig.~\ref{fig:distillation_repetition code} illustrates the quantum circuit for distillation by the $[3,1,3]$ repetition code with a parity-check matrix
\begin{equation}
H_D = \begin{bmatrix}
%\begin{array}{c}
1& 1&  0\\
1&0&1
%\end{array}
\end{bmatrix} ,
\label{eq:repetition_parity}
\end{equation}
for the ancilla state $\ket{0}_L$ of an $n$-qubit quantum CSS code.
We now demonstrate the above protocol by distilling the ancillas of the $[[7,1,3]]$ Steane code \cite{Ste96}, with stabilizer generators
\begin{align*}
g_1&=Z_1Z_4Z_5Z_7,\\
g_2&=Z_2Z_4Z_6Z_7,\\
g_3&=Z_3Z_5Z_6Z_7,\\
g_4&=X_1X_4X_5X_7,\\
g_5&=X_2X_4X_6X_7,\\
g_6&=X_3X_5X_6X_7,
\end{align*}
and logical operators $\bar{X}=X_1X_2X_4$, $\bar{Z}=Z_1Z_2Z_4$.
%{We will carry out the protocol using the 3-bit classical repetition code whose parity-check matrix is given in Eq.~(\ref{eq:repetition_parity}).}  We first demonstrate this distillation protocol with an example in detail.

\rmark{Suppose the $[3,1,3]$ repetition code is used for distilling several noisy codewords of the Steane code.
First, the noisy codewords are divided in to groups of $m=3$. Consider one group with the three noisy Steane codewords $E_1\ket{0}_L,$ $E_2\ket{0}_L, $ and $E_3\ket{0}_L$  prepared independently with errors $E_{1,2,3}$, where $E_1=X_1X_2X_4=\bar{X}$, $E_2=X_3$, and $E_3=X_6X_7$.}  Note that $E_1$ and $E_3$ are uncorrectable errors for the Steane code.  The $X$ error syndromes are
\[
\begin{array}{cc|c}
E_1:& 000&1 \\
E_2:&001&0 \\
E_3:&110&0\\
\end{array}
\]
with respect to $g_1,g_2,g_3,$ and $\bar{Z}$, respectively.
After the (perfect) distillation circuit $U_D$ by the $[3,1,3]$ code (Fig.~\ref{fig:distillation_repetition code}), the errors become
$E_1'=E_1$, $E_2'= E_1 E_2= X_1X_2X_3X_4$, and $E_3'=E_1E_3=X_1X_2X_4X_5X_6$.
Then measuring bitwise the second and the third codewords, and calculating the parities of $g_1,g_2,g_3$ and $\bar{Z}$,
we have their syndrome bits
\[
\begin{array}{cc|c}
\sigma^{\tiny (1)}:&001&1,\\
\sigma^{\tiny (2)}:&110&1.
\end{array}
\]
Now we can use the parity check matrix of the $[3,1,3]$ repetition code to recover the four syndrome bits of the first codeword:
\[\begin{array}{cc|c}
\tilde{s}^{\tiny (1)}:& 000&1.\end{array}
\]
Since the fourth bit is $1$, we apply logical operator $\bar{X}$ to the first codeword to correct the logical error and the final state is $\ket{0}_L$.  Thus we have fault-tolerantly prepared an ancilla $\ket{0}_L$.
\eep

\rmark{Now we carry out protocol I  for general CSS codes as follows.}
Suppose the $X$ errors on the $m$ ancillas are $X_{e^{\tiny (1)}}$, $\dots,$ $X_{e^{\tiny (m)}}$, where $e^{\tiny (j)}= e_{1}^{\tiny (j)}\cdots e^{\tiny (j)}_{n}\in \mathbb{Z}_2^n$.  Let $s^{\tiny (1)}=e^{\tiny (1)}\sfH_Z^T,\dots,s^{\tiny (m)}=e^{\tiny (m)}\sfH_Z^T$.  The decoding operator $U_D$ will transform $X_{e_{j}^{\tiny (1)}}\otimes X_{e_{j}^{\tiny (2)}}\otimes \cdots \otimes X_{e_{j}^{\tiny (m)}}$ to
\[
X_{e_{j}^{\tiny (1)}}\otimes  \cdots \otimes X_{e_{j}^{\tiny (k)}}\ot X_{s_X^j}
\]
for $j=1,\dots, n$, where $s_X^j\in \mathbb{Z}_2^r$ is the error syndrome of $e_{j}^{\tiny (1)}\cdots e^{\tiny (m)}_{j}\in \mathbb{Z}_2^m$ with respect to $\sfH_D$.  If we could measure $s_X^j$,  we could decode $e_j^{i}$ directly and then obtain $k$ clean ancillas.

However, the measurement outcomes $\nu^{\tiny (1)}, \dots, \nu^{\tiny (r)}$ of step 3 are actually a disturbed version of $s_X^j$:
%\begin{align}
%\left(\begin{array}{c}\nu^{\tiny (1)}\\ \vdots\\ \nu^{\tiny (r)}\end{array}\right) =&
%\left( {s_X^{1}}^T \cdots {s_X^{n}}^T\right)
%+ \left(\begin{array}{c} c_1\\\vdots\\c_r \end{array}\right), \notag\\
%=&\sfH_D \begin{pmatrix} {e^{\tiny (1)}}  \\\vdots \\ {e^{\tiny (m)}}  \end{pmatrix}
%+ \left(\begin{array}{c} c_1\\\vdots\\c_r \end{array}\right),
%\label{eq:outcome}
%\end{align}
\begin{align}
\left({\nu^{\tiny (1)}}^T \cdots {\nu^{\tiny (r)}}^T\right) =&
\left(\begin{array}{c} {s_X^{1}}\\ \vdots\\ {s_X^{n}}\end{array}\right)
+ \left( c_1^T \cdots c_r^T \right), \notag\\
=& \left( {e^{\tiny (1)}}^T  \cdots  {e^{\tiny (m)}}^T \right)\sfH_D^T
+ \left( c_1^T \cdots c_r^T \right),
\label{eq:outcome}
\end{align}
where the row vectors $\{c_j\}$ are unknown codewords of the classical code with parity check matrix $\sfH_Z$.  Since we do not know the codewords, we cannot learn $s_X^j$.

On the other hand, we still \emph{can} learn the quantum error syndrome
\[
s^{\tiny (1)}=e^{\tiny (1)}\sfH_Z^T,\dots,s^{\tiny (m)}=e^{\tiny (m)}\sfH_Z^T
\]
from $\nu^{\tiny (1)}, \dots, \nu^{\tiny (r)}$.  Multiplying (\ref{eq:outcome}) from the right by $\sfH_Z^T$,  we have
\begin{align}
 \sfH_D \begin{bmatrix} {s^{\tiny (1)}}  \\\vdots \\ {s^{\tiny (m)}}  \end{bmatrix}= \begin{bmatrix} \nu^{\tiny (1)}\sfH_Z^T \\\vdots \\  \nu^{\tiny (r)}\sfH_Z^T  \end{bmatrix}
 \triangleq \begin{bmatrix} \sigma^{\tiny (1)} \\\vdots \\  \sigma^{\tiny (r)} \end{bmatrix}.
\end{align}

Then we can choose any decoder of $\cC_D$ to recover the~$k$ error syndromes $\tilde{s}^{\tiny (i)}$ of the target ancillas as in step~4.  That is, we are retrieving a particular syndrome bit~$j$ of every ancilla $\tilde{s}^{\tiny (1)}_j, \dots, \tilde{s}^{\tiny (m)}_j$ in one classical decoding procedure.  From that, we can correct all the $X$ errors in the target $k$ ancillas (and also any logical $X$ errors if we are distilling $\ket{0}_L$).
%\noindent\textbf{Remark:}
\begin{remark}
As long as fewer than $\lfloor \frac{d-1}{2}\rfloor$ of the $m$ syndrome bits $s^{\tiny (1)}_j, \dots, s^{\tiny (m)}_j$ are 1s, they can be recovered by the classical decoding, assuming that the quantum gates in the distillation circuit are perfect.
\end{remark}

\begin{remark}
The distillation protocol depends only on the error-correcting power of the classical code $\cC_D$, and not on the error-correcting ability of the stabilizer states being distilled.  If $n=1$, this procedure reduces to the standard classical decoding of~$\cC_D$.
\end{remark}

After round 1, the remaining $k$ ancillas from the group will have lower rates of $X$ errors than they started with.  However, $Z$ errors on the parity-check ancillas can propagate via the CNOTs back onto these $k$ ancillas, increasing the rate of $Z$ errors (and also correlating the errors across the ancillas).  How do these changes compare?  Assume that the original rates of $Z$ and $X$ errors are both $p$.  The rate of $Z$ errors on the remaining $k$ ancillas will increase to $\sim (\beta+1)p$, where $\beta \leq r$ is the number of parity checks that each qubit is included in (or the number of 1s in each row of~$\sfA$).  The rate of $X$ errors goes from $p$ to $cp^{t+1}$, where $c$ is a constant that depends on the details of the codes.  If $p$ is not too big, the rate of $Z$ errors has grown roughly by a constant factor, while the rate of $X$ errors has been substantially reduced.

\begin{figure}[h]
\includegraphics[scale=0.42]{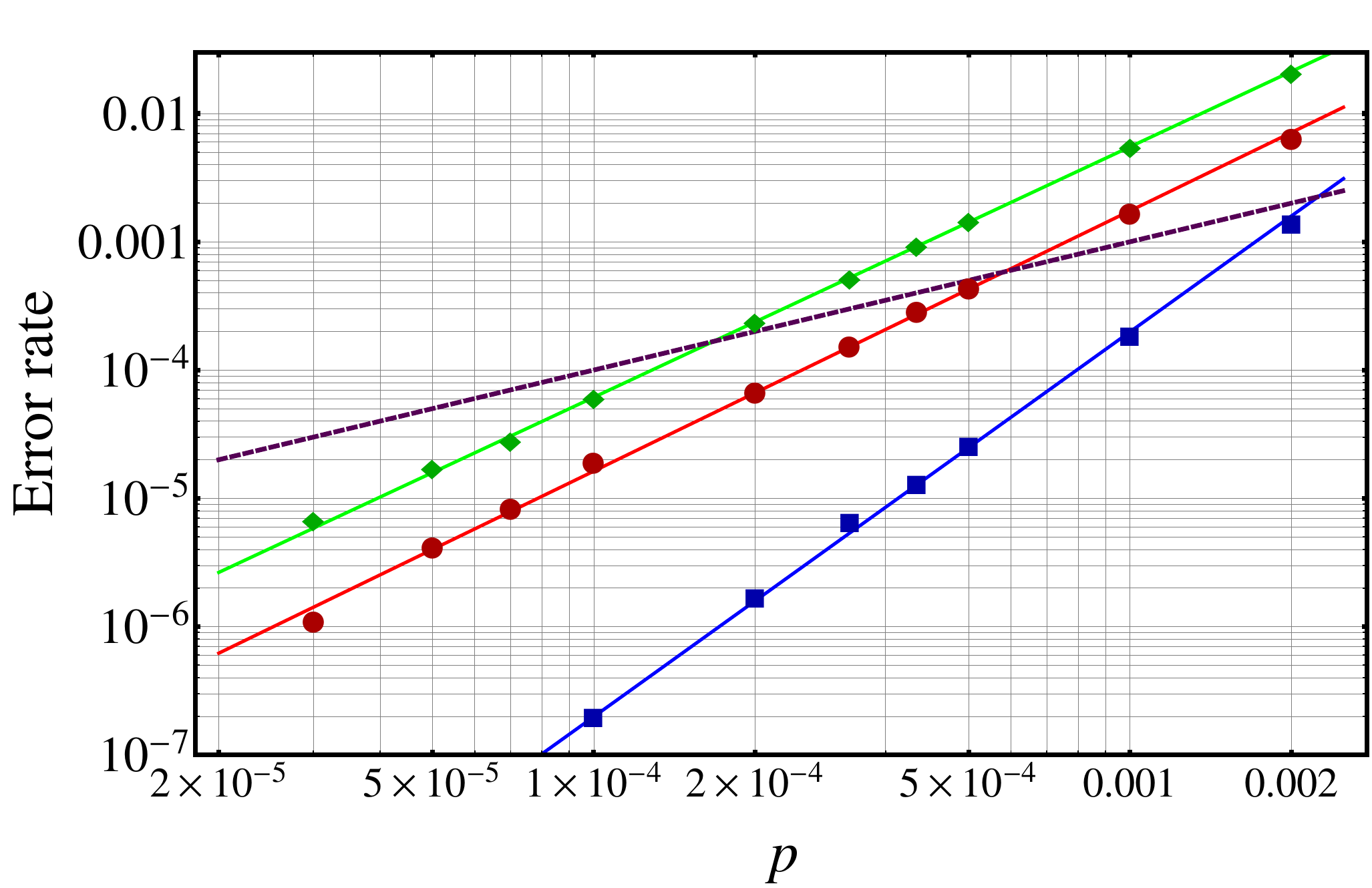}
\caption{(Color online.)  
Ancilla distillation by 1) the $[7,4,3]$ Hamming code (green diamonds); 2) the $[3,1,3]$ repetition code (red dots); 3) the $[5,1,5]$ repetition code (blue squares). The dashed line is the rate without distillation. Up to  $7\times 10^{8}$ iterations are used for each point.}
\label{fig:distillation_steane_code}
\end{figure}

After round 2, we are left with a fraction $(\frac{k}{m})^2$ of our original ancillas.  The rate of $Z$ errors will go from $(\beta+1)p$ to $c((\beta+1)p)^{t+1} = c' p^{t+1}$, and the rate of $X$ errors will go from $c p^{t+1}$ to $(\beta+1)c p^{t+1} = c'' p^{t+1}$. {(So the rate of an arbitrary Pauli error is roughly $\tilde{c}p^{t+1}$ for some $\tilde{c}$.)}  These constants will not generally be equal to each other (and indeed, the starting rates for $X$ and $Z$ might not have been equal ); one might use two different classical error-correcting codes in the two rounds so that the final rates both end up below some desired fraction.
{To reach a desired target error rate, this procedure could be iterated;} or one could just vary
the distances of the {classical} codes used depending on the original error rates and the desired final error rates.

\be
{In addition to the $[3,1,3]$ repetition code above, we simulated distillation by the $[5,1,5]$  repetition code and the $[7,4,3]$ Hamming code \cite{MS77}.  The results are shown in Fig.~\ref{fig:distillation_steane_code}.  The simulations begin with preparation of noisy ancillas by the CSS encoding circuit.  We assume that during the ancilla preparation each individual qubit suffers independent depolarizing errors with rate $p$.  Noisy CNOTs in the encoding circuit are modeled as a perfect CNOT, followed by no error (with probability $1-p$) or one of the nonidentity two-fold Paul operators (e.g., $X\ot Y$, $I\ot Z$, $X\ot I$, $\ldots$) with equal probabilities $p/15$.  For our initial analysis, we assume that the distillation circuit does not itself contain errors; the issue of imperfect distillation circuits will be addressed in the discussion section.}

{We define the final error rate to be the probability of any Pauli error left in the target ancillas after \emph{two rounds} of distillation.  This estimate is \emph{pessimistic}, since it ignores the likelihood that the residual errors are correctable in the next error correction cycle.}

After the noisy encoding circuit, the error rate on each qubit will increase to $\sim (\alpha+1)p$, where $\alpha$ is the number of stabilizer generators each qubit is involved.  If $p$ is small enough, the distillation protocol will work. As can be seen in the logarithmic plot in Fig.~\ref{fig:distillation_steane_code}, each curve appears linear with slope $t+1$ and the ``threshold" for each code is specified by $\log p_{\text{th}}= -\frac{1}{t}\log \tilde{c}$.  (By ``threshold,'' we mean the crossing point of a curve with the dashed line. That is, the point where the distillation error rate and the physical rate $p$ are equal.)  {Thus, we say that a stabilizer state can be fault-tolerantly prepared if the error rate is below the threshold for the classical code used in distillation.}
\eep
%\bmark{
%For these simple examples, a code with high distance and low code rate would have better performance at the expense of additional CNOTs and classical computing power.}
\bmark{For these simple examples, we can see that the [5,1,5] code with higher distance has asymptotically better behavior in the low physical error rate regime but the yield rate is low. On the other hand, the [7,4,3] code has a larger yield rate but worse performance.}
%So classical codes with both high code rate and distance are expected in this scenario.}
 Since there are many efficient classical codes with high rate and good error-correcting ability, we can certainly find better candidates for the protocol with extra cost.

Finally we briefly discuss candidates for large quantum codes whose raw ancilla state preparation by noisy Clifford circuits will have sufficiently low error rates.  It is known that quantum CSS codes built from classical doubly-even codes allow transversal Clifford gates~\cite{Shor96,Ste99N}:  such codes as the Steane code, the $[[23,1,7]]$ quantum Golay code, or other quantum quadratic-residue codes \cite{LL11}.  We can concatenate a large quantum block code with, for example, the quantum Golay code, whose Clifford operations can be done transversally, and these ancillas can be prepared fault-tolerantly~\cite{PR12}. With quantum Golay code blocks at the bottom level, the Clifford encoding circuit for the large quantum code can be transversally implemented, and quantum error correction by the quantum Golay code can be inserted into the circuit, if necessary.  Therefore, the error rate of the raw ancillas can be suppressed to below the ``threshold" of distillation.  If we constantly apply quantum error correction by the Golay code, the output should be very good, but the overhead will be large.

{Another intriguing possibility is to perform distillation steps at several points during the encoding circuit for the ancillas, to remove errors before they can spread widely.  Efficient distillation would require the use of very high-rate classical codes.}

\subsection{Distillation by Quantum CSS Codes} \label{sec:quantum distillation}

Instead of using classical codes in two steps to correct both $X$ and $Z$ errors, we can similarly use an $[[m,k]]$ quantum CSS code $\cQ_D$ to distill the desired ancillas of an $[[n,1]]$ code.  For simplicity, suppose $\cQ_D$ is defined by $\sfH_Z'$ and $\sfH_X'$ of dimension $r_Z'\times m$ and $r_X'\times m$ matrices  $(k=m-(r_Z'+r_X')>0)$ with full rank $r_Z'$ and $r_X'$ and $\sfH_X' \sfH_Z'^T =0.$  $\cQ_D$ has $r_Z'$ $Z$ generators, and $r_X'$ $X$ generators.  The check matrix of an $[[m,k]]$ CSS code can be written in the following form~\cite{NC00}:
\[
\begin{pmatrix}\sfH_Z'& 0\\
0&\sfH_X'
\end{pmatrix} =   \left(
\begin{array}{ccc|ccc}   \sfI_{r_Z'}& \sfA& \sfB &  0 & 0  &   0\\
                                 0 &  0& 0 &  \sfD & \sfI_{r_X'}&  \sfF\\
                                 \end{array}\right),
                                 \]
where  $\sfA$, $\sfB$, $\sfD$, and $\sfF$ are binary matrices of appropriate dimensions.  {We use a particular encoding circuit as  follows:} to the $k$ information qubit state $\ket{\phi}$, append $r_Z'$ ancilla qubits in the state $\ket{0}$ and $r_X'$ ancilla qubits in the state $\ket{+}$, which will correspond to the $r_Z'$ $Z$ generators and $r_X'$ $X$ generators, respectively, after encoding.
That is, we apply an encoding circuit to the initial state
\[
\ket{0}^{\otimes r_Z'}\otimes \ket{+}^{\otimes r_X'}\otimes \ket{\phi},
\]
which corresponds to a check matrix
\[
\left(\begin{array}{ccc|ccc}\sfI_{r_Z'}&  0 & 0 & 0 & 0 & 0\\
          0 & 0&  0 &  0 &\sfI_{r_X'}& 0\\
          \end{array}\right).
\]
The encoding circuit $U'_D$, which is a unitary operator, will then consist only of CNOT gates~\cite{Gra02}.
It can be verified that this works for any CSS codes.
 and we postpone this justification to Appendix~\ref{sec:encoding justification}.

Suppose an error operator $E\in \cG_m$ occurs on a codeword of $\cQ_D$.  We decode by running the above encoding circuit backwards, so that the error syndrome information will be contained in the ancilla qubits.  After the decoding circuit ${U'_D}^\dag$, we have the following transformed error operating on the initial state $\ket{0}^{\otimes r_Z'}\otimes \ket{+}^{\otimes r_X'}\otimes \ket{\phi}$:
\[
{U'_D}^{\dag} E U'_D \triangleq X_{s_X} \otimes Z_{s_Z} \otimes L_E,
\]
where $s_X\in \mathbb{Z}_2^{r_Z'},$ $s_Z\in \mathbb{Z}_2^{r_X'}$ are the syndrome vectors (hence $X_{s_X}\in \cG_{r_Z'}$, $Z_{s_Z}\in \cG_{r_X'}$) and $L_E\in \cG_k$ is an logical error operator.

We then measure the first $r_Z'$ ancillas in the $Z$ basis, which gives the $X$ error syndrome $s_X$,
and the other $r_X'$ ancillas in the $X$ basis, which gives the $Z$ error syndrome $s_Z$.  From that, we can figure out what corrections, if any, need to be applied to the $k$ information qubits.  This leads to the following distillation scheme:
\medskip

\noindent\emph{Distillation Protocol II:}

Suppose we want to distill a CSS stabilizer state of the $[[n,1]]$ code $\cQ$ defined at the beginning of this section.
\begin{enumerate}[1)]
\item Start with $m$ imperfect copies of a stabilizer state of $\cQ$.
Do a transversal ${U'_D}^\dag$   on the $m$ copies of the $n$-qubit system.

\item For the $r_Z'$ (respectively  $r_X'$) systems in the positions corresponding to the $Z$ (resp.  $X$) ancillas, we measure all the qubits in the $Z$ (resp.  $X$) basis and let  $\nu^{\tiny (1)},\dots, \nu^{\tiny (r_Z')}\in \mZ_2^m$ (resp. $\nu^{\tiny (1)},\dots, \nu^{\tiny (r_X')}\in \mZ_2^m$) be the outcomes.

\item Calculate $\sigma_X^{\tiny (i)}\triangleq \nu_Z^{\tiny (i)}  \sfH_Z^T$ for $i=1,\dots,r'_Z$.  For $j=1, \dots, r_Z$, use $[{\sigma_{X}^{\tiny (1)}}]_j \cdots [{\sigma_X^{\tiny (r_Z')}}]_j$ as a classical error syndrome with respect to the parity-check matrix $\sfH_Z'$ and use a corresponding decoder to find the most probable error vector $\tilde{s}^{\tiny (1)}_j \cdots \tilde{s}^{\tiny (m)}_j$ with this error syndrome.  Then $\tilde{s}^{\tiny (i)}_1\cdots \tilde{s}^{\tiny (i)}_{r_Z}$ is the estimated $X$ error syndrome of the $i$th target ancilla.  We can correct the $X$ errors, if any,  (or just keep track of them).

\item Repeat 3) but with $X$ and $Z$ switched everywhere.

\item If  the distillation target is $\ket{0}_L$ or $\ket{1}_L$ (resp. $\ket{+}_L$ or $\ket{-}_L$), then calculate the parity of $\bar{Z}$ (resp. $\bar{X}$) from $\nu_Z^{\tiny (i)}$  (resp. $\nu_X^{\tiny (i)}$) and estimate the syndrome bits for the logical operators as in the previous step.  Correct the logical error $\bar{X}$ (resp. $\bar{Z}$), if any,  (or just keep track of them).

\hfill$\square$
\end{enumerate}

This protocol is very similar to Distillation Protocol~I. Thus we omit the explanation of how it works.  We end up with $k$ much cleaner copies of the $n$-qubit CSS stabilizer states of~$\cQ$.  Again, this procedure could be iterated, or we can use a good enough $[[m,k]]$ code.  This is like a concatenation of the $n$-qubit code $\cQ$ with the $[[m,k]]$ code $\cQ_D$, but only the decoding on the second level, $\cQ_D$, is applied.

Protocol~I by classical codes is more flexible than Protocol~II, since any classical codes can be used in Protocol~I and the codes can be different in two rounds of distillation; whereas only dual-containing codes can be applied in Protocol~II.  On the other hand, dual-containing codes used in Protocol~II can also be applied in Protocol I, and the number of CNOTs for correcting both $X$ and $Z$ errors are roughly the same in the two protocols.  Basically, the performance of these two protocols are strongly related to the performance of the classical codes, so we omit simulations of Protocol~II here.

\section{Ancilla saving}
\label{sec:ancilla saving}

Clean ancillas $\ket{0}_L$ and $\ket{+}_L$ in Steane syndrome extraction are expensive resources.  We would like to {use as few of them as possible} during syndrome measurement, as long as errors do not accumulate seriously.  In the following, we show that classical coding can also reduce ancilla consumption in Steane syndrome extraction; it turns out that this problem is equivalent to the distillation problem in Sec.~\ref{sec:classical distillation}.  We assume that the quantum circuits for error correction are perfect, and the ancillas are assumed to be \emph{clean} in the following discussion.

Suppose we have $m$ codewords  $\ket{\psi_1},\dots, \ket{\psi_m}$ of the $[[n,1]]$ CSS code $\cQ$ defined by $\sfH_Z$ and $\sfH_X$.  Let  $X_{e^{\tiny (j)}}Z_{f^{\tiny (j)}}$ be the error corrupting  $\ket{\psi_j}$ for $j=1,\dots, m$.  If Steane syndrome extraction is used, $m$ (perfect) ancillas $\ket{+}_L$ are required to measure the $X$ error syndromes $e^{\tiny (j)}\sfH_Z^T$.  Our goal here is to estimate the $X$ error syndromes  by using only $r$ $(<m)$ clean ancillas $\ket{+}_L$.  The treatment  for  $Z$ error is similar.

Let $\cC_D$ be an $[m,k=m-r,d]$  classical code   with  parity-check matrix $\sfH_D= [  \sfA^T \ \sfI_r ]$.
Assume we have  $r$ clean ancillas $\ket{+}_L$ and $r$  clean ancillas $\ket{0}_L$.
\medskip

\noindent\emph{{Ancilla Saving Protocol}:}
\begin{enumerate}[1)]
\item Apply transversal CNOTs from $\ket{\psi_j}$ to the $j$-th $\ket{+}_L$ for $j=k+1,\dots, m$.

\item Apply a transversal CNOT from $\ket{\psi_{i}}$ to the $j$-th $\ket{+}_L$ if  $[\sfA]_{i,j}=1$.

\item Do steps 3 and 4 of Distillation Protocol~I.

\item Apply transversal CNOTs from the $j$-th $\ket{0}_L$  to $\ket{\psi_j}$ for $j=k+1,\dots, m$.

\item Apply a transversal CNOT from the $j$-th $\ket{0}_L$  to $\ket{\psi_{i}}$ to if  $[\sfA]_{i,j}=1$.

\item Do steps 3 and 4 of Distillation Protocol~I, but with $X$ and $Z$ switched everywhere.

\hfill$\square$
\end{enumerate}

\begin{figure}
\centering
\includegraphics[scale=0.3]{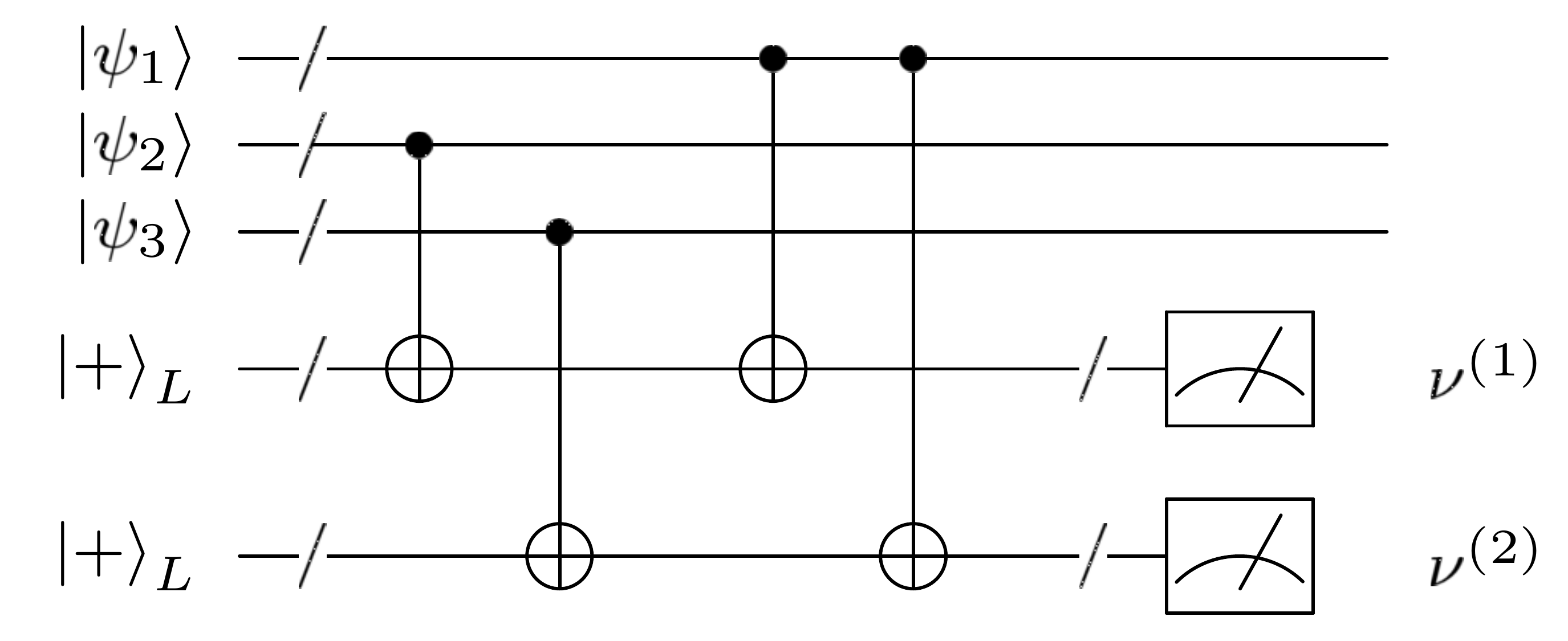}
\vspace{-0.5cm}
\caption{The circuit for syndrome measurement of three data blocks with two ancilla blocks by the $[3,1,3]$ repetition code. }
\label{fig:ancilla saving}
\end{figure}

Fig. \ref{fig:ancilla saving} demonstrates the circuit for the $X$ syndrome extraction of three data blocks with two ancilla blocks by the $[3,1,3]$ repetition code.  Observe that this circuit is essentially equivalent to the circuit in Fig.~\ref{fig:distillation_repetition code}.  We can combine the $X$ errors from the second and third encoded states with the two clean ancillas $\ket{+}_L$, respectively, and then remove those two encoded states.  As a consequence, our ancilla saving protocol is equivalent to the distillation protocol by classical codes.

We may compare the error correction performance of this ancilla saving protocol by $\cC_D$ on codewords of $\cQ$, say $\cQ+\cC_D$, with the original Steane extraction scheme.  A good figure of merit for comparison is the channel fidelity of a quantum code over a noise channel \cite{Schumacher96,RW06}.  For simplicity, here we assume the independent single-qubit noise channel is
\[
\mathcal{E}(\rho)= (1-p)\rho + p X\rho X ,
\]
for any single-qubit state $\rho$.  Unquestionably the channel fidelity is expected to drop with fewer ancillas in the saving protocol, and both the encoding and decoding complexities will increase.  Thus, we have a tradeoff between channel fidelity, gate complexity, and ancilla consumption.

\be
Consider the $[[7,1,3]]$ Steane code.  It is known that the channel fidelity of a quantum code over a Pauli channel is the probability of correctable errors \cite{LB12}:  that is, the probability of a set of coset leaders and their degenerate errors.  It is more complicated to calculate the channel fidelity of the saving protocol.  We estimate it as follows:

Let $F_C(\cE)_i$ be the channel fidelity of sending $\ket{\psi_i}$ through $\cE^{\otimes n}$ for $i=1,\dots, m$.  Then the average channel fidelity is
\[
\overline{F_C}(\cE)= \frac{1}{m}\sum_i F_C(\cE)_i.
\]
Since $\ket{\psi_i}$ are correlated in the saving protocol, the previous result of $F_{C}(\cE)$ cannot be applied here.
Thus we apply Monte Carlo methods to approximate $\overline{F_C}(\cE)$:

{
\begin{enumerate}[1)]
\item Fix $p$. Set  $i:=1$.

\item Apply Pauli $X$ errors  $X_{e^{\tiny (1)}}^{i},\dots, X_{e^{\tiny (m)}}^{i}$ to perfect information states $\ket{\psi_1},\dots, \ket{\psi_m}$, where $X_{e^{\tiny (j)}}^{i}\in \cP_n$ is an $n$-fold Pauli $X$ error, randomly  generated according to the probability distribution of $\cE$.

\item Use the ancilla-saving code $\cC$ to recover the error syndromes $s_1,\dots, s_m$ for $X_{e^{\tiny (1)}}^{i},\dots,X_{e^{\tiny (m)}}^{i}$, respectively.  If there is no logical error on $\ket{\psi_j}$ after decoding, $X_{e^{\tiny (j)}}^{i}$ is correctable. Set $i:=i+1$.

\item Repeat steps 2 and 3 $N$ times.

\item Count the number of correctable errors in $\{X_{e^{\tiny (j)}}^{i}\}$, say $M$.  Then $\overline{F_C}(\cE)$ is approximated by $\frac{M}{mN}$.
\end{enumerate}
}

\begin{figure}[htb]
\centering
\hspace{0.1cm}\includegraphics[scale=0.45]{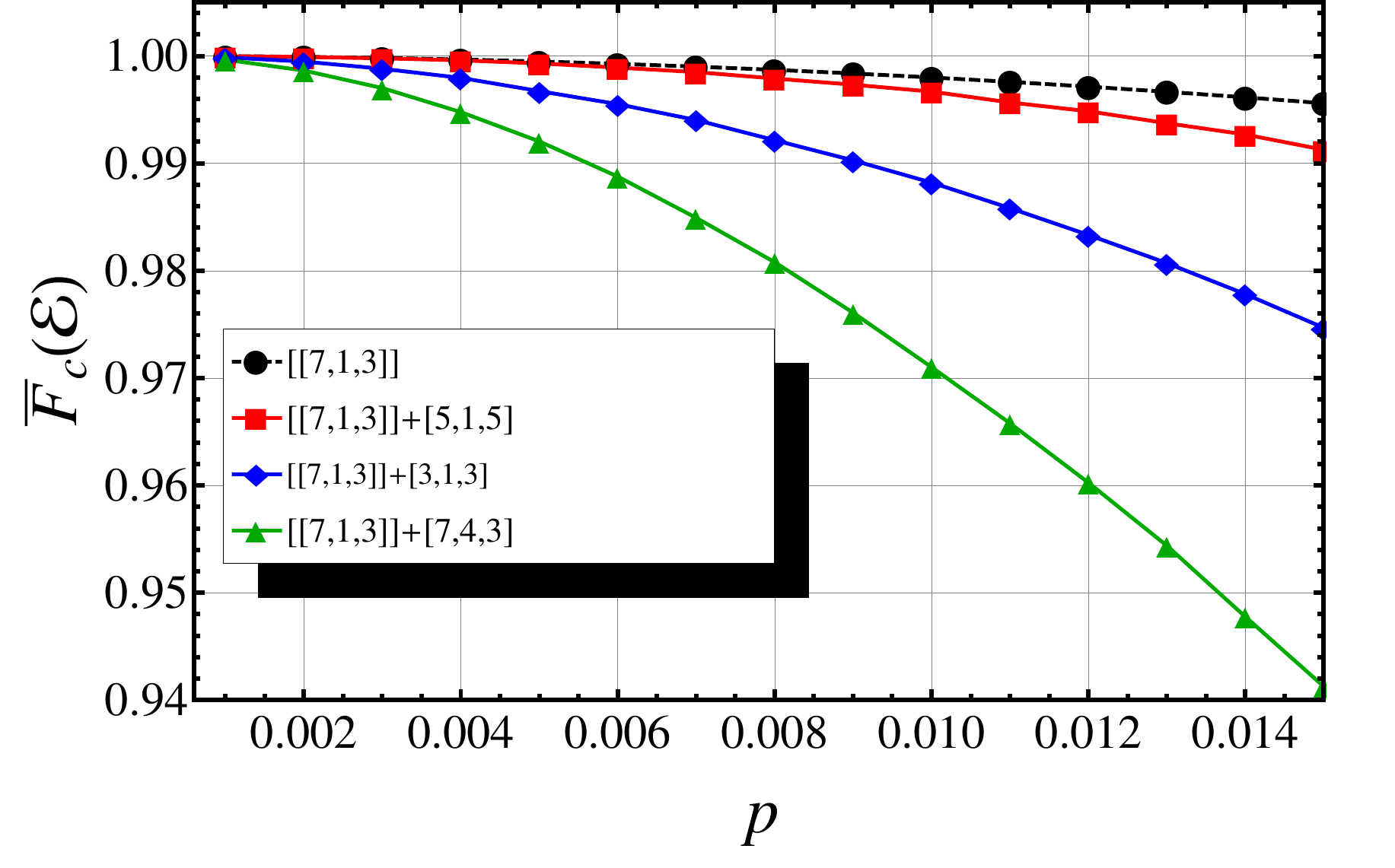}\\%(\text{a})&(\text{b})\\
\caption{(Color online.)  The average channel fidelity of the Steane code without or with ancilla saving by the $[3,1,3]$, $[7,4,3]$, or  $[5,1,5]$ code. The number of iterations for each point is up to $7\times 10^{8}$.}
\label{fig:ancilla_saving_fidelity}
\vspace{-0.2cm}
\end{figure}

Fig.~\ref{fig:ancilla_saving_fidelity} illustrates plots on $\overline{F_C}(\cE)$ for various ancilla saving protocols:  the $[7,4,3]$ Hamming code, and the $[3,1,3]$ and $[5,1,5]$ repetition codes, together with $F_{C}(\cE)$ without ancilla-saving.  As shown in Fig. \ref{fig:ancilla_saving_fidelity}, at $p=0.01$, the channel fidelity for  $[[7,1,3]]+[3,1,3]$ drops from 0.998 to {0.988}.  Since only two ancilla states are used, we reduce the ancilla consumption by $33.3\%$  by using one additional  transversal CNOT and increased classical computing complexity.  (Note that the circuit for preparing a $\ket{+}_L$ has nine CNOTs.)  For  the $[7,4,3]$ code, the fidelity drops significantly to save $\frac{4}{7}=57.1\%$ ancillas.  For the $[5,1,5]$ code, the fidelity drop at $p=0.01$ is less than {$0.2\%$},  while  $\frac{1}{5}=20\%$ of the ancillas are saved.

\eep

\begin{figure}[htb]
\centering
\includegraphics[scale=0.45]{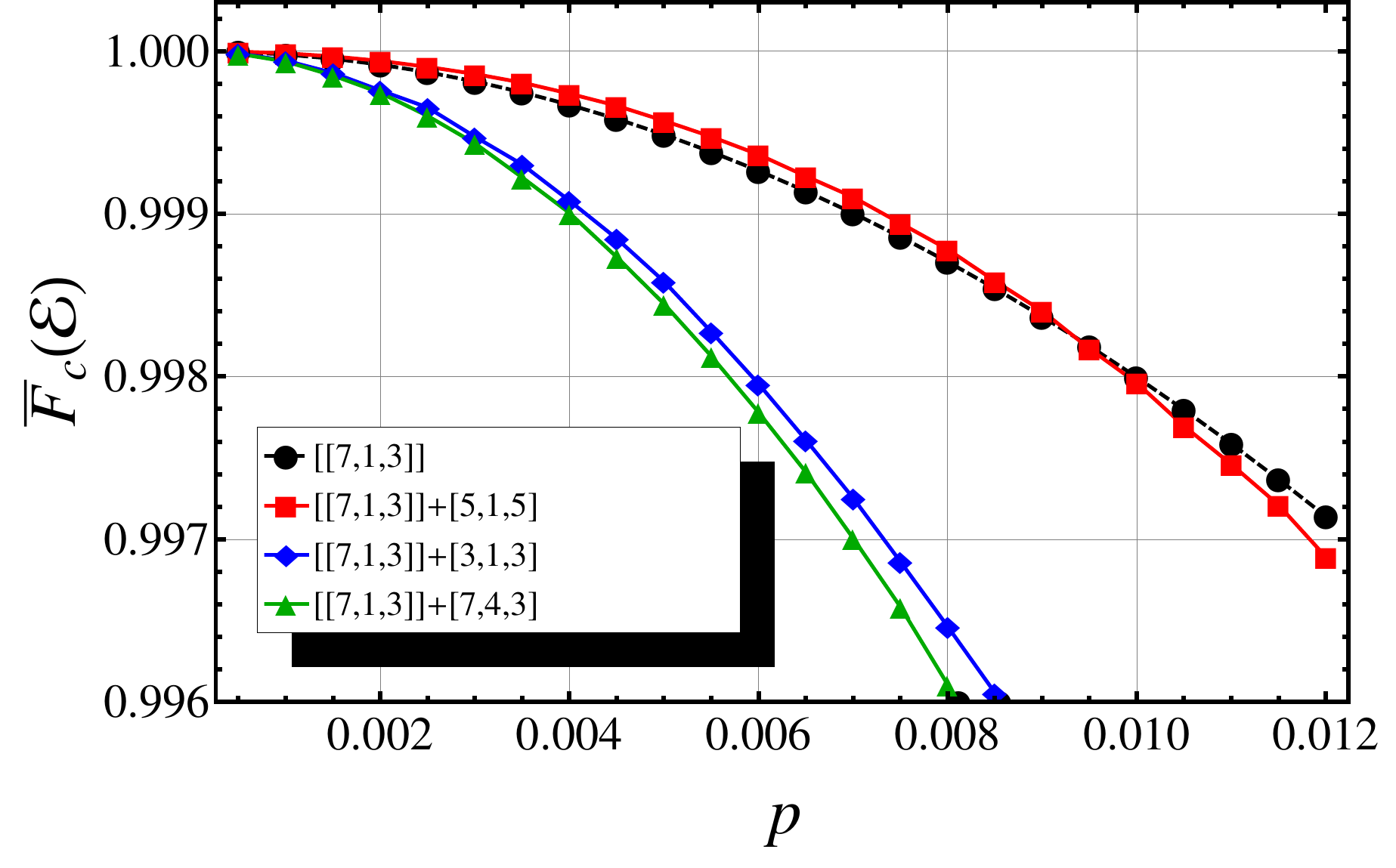}
\caption{(Color online.)  The average channel fidelity of Steane code with ancilla saving by the $[3,1,3]$, $[7,4,3]$, or  $[5,1,5]$ code at \rmark{physical error rate} $p$.  The number of iterations for each point is up to $7\times 10^{8}$.}
\label{fig:ancilla_saving_fidelity effective}
\end{figure}

When the  ancilla consumption rate is fixed, we can increase the frequency of quantum error correction with the ancilla saving protocol, which is equivalent to lowering the error rate on the data qubits.
\rmark{Let us define the effective error rate as the (accumulated) physical error rate between two error correction steps.
Here the effective error rate of an $[[n,1]]$ code without ancilla saving is simply called the \emph{physical error rate}.
If the physical error rate is $p$,
then the effective error rate of an $[[n,1]]+[m,m-r]$ ancilla saving protocol is $rp/m$,
assuming that quantum error correction is sufficiently fast.  Let $F_{\text{o}}^p$ and $F_{\text{comb}}^{p}$ be the channel fidelities of the original and the $[[n,1]]+[m,m-r]$ protocols at \emph{effective} error rate $p$, respectively.
Apparently we have $F_{\text{o}}^0=F_{\text{comb}}^{0}=1$ and $F_{\text{o}}^p > F_{\text{comb}}^{p}$ for $p>0$.
 Thus by the continuity of fidelity,  there exists $p^*$  such that $F_\text{o}^{p^*}= F_{\text{comb}}^{rp^*/m}$.
Here we define the \emph{effective channel fidelity} of an $[[n,1]]+[m,m-r]$ protocol as the average channel fidelity of the protocol at effective error rate $rp/m$.
 Hence for physical error rate $p< p^*$ the effective channel fidelity of the $[[n,1]]+[m,m-r]$ protocol is higher.}
Let us consider the above example again.  Fig.~\ref{fig:ancilla_saving_fidelity effective} plots these channel fidelities. %at physical error rates relative to the original $[[n,1]]$ protocol.
As can be seen, applying the ancilla saving protocol with the $[5,1,5]$ code is better than the original scheme for $p<0.00925$, but there is no fidelity gain for the other two codes. Of course, this fidelity gain was at the cost of some additional CNOT gates and classical decoding steps.

\section{Discussion}
\label{sec:discussion}

A straightforward approach to achieving FTQC is to implement logical gates transversally, but no quantum stabilizer code can have a universal gate set of quantum computation that can be transversally implemented \cite{EK09,ZCC11}.  As a consequence, universality is usually accomplished by the assistance of certain ancillas, prepared by {magic state distillation} \cite{BK05}.  A large literature is devoted to reducing the overhead of magic state distillation~\cite{SJ12,Eastin13,Jones13a,Jones13b,Jones13c,PR13}, since it  may dominate the overall resources needed for FTQC~\cite{FMMC12}.

{The current work is motivated by a different approach to fault-tolerance:  the use of a set of CSS codes to store and process quantum information.  By teleporting logical qubits between code blocks that admit different sets of transversal gates, it is possible to perform a universal set of logical gates \cite{BZHJL14}.  We have illustrated distillation with the simple example of the Steane $[[7,1,3]]$ code, but the procedure can readily be generalized to multi-qubit CSS codes.  In particular, they can distill any CSS stabilizer state which is the simultaneous +1 eigenstate of all the stabilizer generators and a set of commuting logical operators, such that each logical operator includes the identity and either only $X$ operators or only $Z$ operators.  This limitation means that it is not possible to distill completely general stabilizer states (or even general CSS stabilizer states) by the methods presented here, though generalizations of this scheme may make that possible.  However, the protocols presented in this paper can distill all the logical ancillas needed for the teleportation-based FTQC scheme of \cite{BZHJL14}.

Given the ability to prepare a set of suitable CSS stabilizer states, only transversal circuits and single-qubit Pauli measurements are needed for FTQC.  In particular, magic states are not needed.  (Other schemes without magic state distillation have also been proposed, such as \cite{PR13,JL14,ADP14,BC15,JTS16}.)  The results of this paper show that in principle this approach to FTQC is possible.  However, the overhead for distillation dominates in this scheme, and we need to further analyze and quantify both the cost of distilling ancillas for various codes, and the performance of distillation in the presence of errors in the distillation circuit.

Our distillation protocol is a combination of the Steane syndrome extraction method and classical coding so that stabilizer states can be fault-tolerantly prepared.  However, more work is needed to show that the overall scheme is both fault-tolerant and efficient enough to be useful at realistic error rates.  In particular, noise in the distillation circuit will have two important effects.  First, residual errors will be left in the ancilla by the noisy distillation circuit.  Second, errors may degrade the performance so that ancilla errors from the encoding circuit may not be fully corrected.  This first source of error will increase the effective error rate in the computation; but because the distillation circuit is transversal, these errors should be independent across the qubits of a single ancilla.  However, the second type of error would be more serious:  correlated errors across an ancilla can dramatically shorten the lifetime of the quantum codes used in the computation.  Thus, careful modeling and numerical simulations are needed to assess and optimize the performance of distillation in the presence of noise.

Some methods may be used to greatly mitigate these potential pitfalls.}  In the distillation protocol by classical coding, error syndromes of the target ancillas are encoded by the coupling CNOTs and then recovered.  If the transversal circuits for distillation are imperfect, the measured parity-check syndromes $ \nu \sfH_Z^T$ are not reliable, which compromises the efficiency of distillation.  However, this can be handled by learning more parities of the stabilizer generators as suggested by the method in \cite{ALB14,ALB16,Fuji14}.  In particular, we can choose another classical code $\cC_3$ to encode the parity checks of $\sfH_Z$ by appending more redundant rows to the parity-check matrix.  By calculating these additional parity checks,  we can use any decoder of  $\cC_3$ to purify the decoding outputs of $\cC_D$ and obtain more reliable error syndromes about the target ancillas.  To further improve the reliability of distillation, we can also use a final postselection.  By filtering out those noisy blocks with distinct syndromes, residual errors of higher weight can be further reduced {at only a small cost in yield for the protocol.

We have begun to study these methods, and have found some preliminary results.}  A simulation of noisy distillation of the $[[23,1,7]]$ quantum Golay code by the $[3,1,3]$ repetition code, followed by the $[23,11,7]$ Golay code with postselection, suggests an overall yield of $10\%$ by our protocol at a physical error rate of about $10^{-4}\sim 10^{-5}$.  Technical details and performance analyses for some candidate classical codes and quantum  codes will be addressed in a forthcoming paper \cite{ZLB15}.

In  \cite{PR12}, an ancilla verification method is proposed for the quantum Golay code.
{This method can be regarded as a special case of our protocol, but our distillation protocol is potentially more efficient.}  In the verification scheme, pairs of blocks are compared to check errors. If any errors are detected in either block, everything is discarded and the process starts over again.  By contrast, the results of each verification (parity checks) are kept track of in our protocol, and we can take advantage of their correlations  by  classical decoding.  Since good classical codes with high rate and efficient decoders exist, we expect our distillation protocol to have higher throughput.

{In the simulations in this paper, we used a minimum distance decoder of the classical code for distillation.  However,  the error rate of each syndrome bit of an ancilla may depend on the Clifford circuit that generates these faulty ancillas.  We may be able to analyze this dependence and employ other techniques to improve the decoding performance.  This is another future research direction.}

Our distillation protocols are similar to magic state distillation, but there is an important distinction:  because these ancillas are stabilizer states, they can be made using only Clifford gates, and (in principle) can be fault-tolerantly verified.  This would suggest better performance here than in magic state distillation, where one cannot improve the quality of the encoded state by measuring it directly.  At the very least, we should be able to do better in this respect:  with magic state distillation, there is a probability of failure at each iteration step, where you have to discard everything; here, if we detect an error in the logical operators, we can correct them.  Also, only certain codes with special properties can be used for magic state distillation; while a broad range of classical error-correcting codes can be applied in our scheme.

In this paper, we also have shown how to recover accurate error syndromes in Steane syndrome extraction using fewer ancillas, at the cost of higher classical decoding complexity and some additional CNOTs, while sacrificing a little channel fidelity.  Since classical computing power is much cheaper, in general, than expensive quantum resources, it makes sense to exploit classical computing to save quantum resources.  The layout of additional transversal CNOTs depends on the chosen classical code and their cost may or may not be comparable to the complexity saved by preparing fewer ancillas.  However, the overall error-correcting power can be increased when the ancilla consumption rate is fixed.  This protocol shares the same structure as ancilla distillation, and should give a net benefit at least in the regime of low error rates.

{To do quantum error correction, we can also use the Shor syndrome extraction~\cite{DS96,Shor96}. (Knill syndrome extraction~\cite{Knill04} is essentially equivalent to the Steane method up to qubit relocations.)  For codes with low-weight stabilizer generators, Shor syndrome extraction may be preferred since it needs only low-weight ancillas---the cat states $(\ket{0}^{\otimes w}+\ket{1}^{\otimes w})/\sqrt{2}$---of size approximately equal to the weights of stabilizer generators.  The cat states are also stabilizer states and  thus could be prepared by our distillation protocols.  We simply mentioned this since there are already methods of verifying the cat states.}

Finally, our distillation protocols by classical error-correcting codes or quantum CSS codes could also be directly applied to the problem of multipartite entanglement purification for CSS stabilizer states.  Conversely, a protocol for multipartite entanglement purification could potentially lead to a distillation protocol of CSS stabilizer states in FTQC if the multipartite constraint is removed.  {In particular, the purification protocols I/II in~\cite{DAB03,ADB05} are a special case of our distillation protocol~I by the $[2,1,2]$ classical error-detecting code with parity-check matrix
\[
\sfH_D= \left(\begin{array}{cc}1&1\end{array}\right) ,
\]
where an ancilla is discarded if the measurement outcomes are nonzero.  However, we can use a general classical error-correcting code to do error recovery in the distillation process, and hence the efficiency can be better.  Distillation protocol~I could be adapted to a hashing protocol for multipartite entanglement purification as in \cite{ADB05,CL04,HDD06}; however, we omit further discussion for the present, since the main topic of this paper is about fault-tolerant quantum computation.}  {Also, the protocol in~\cite{GKV06} is very similar to our distillation protocol~II.  Since they did not consider the problem in the fault-tolerant scenario, the eigenvalues of the  stabilizers are measured directly in their protocol.  By contrast, we use a transversal decoding circuit and bitwise qubit measurements to recover the eigenvalues of the stabilizers  in our protocol.}

\begin{acknowledgments}
%\emph{Acknowledgements.}--
We thank Ben Reichardt for useful discussions. We also thank Scott Glancy for helpful comments.  This work was supported in part by the IARPA QCS program; by HRL subcontract No. 1144-400707-DS; and by NSF Grant No. CCF-1421078.  This work was also partially funded by the Singapore Ministry of Education (Tier-1 funding), through Yale-NUS Internal Grant IG14-LR001.  T.A.B. also acknowledges funding as an IBM Einstein Fellow at the Institute for Advanced Study.
%Numerical calculation used resources of the Oak Ridge Leadership Computing Facility at the Oak Ridge National Laboratory, which is supported by DOE contract No. DE-AC05-00OR22725.
\end{acknowledgments}

\appendix
\section{Justification of the Encoding for CSS Codes}
\label{sec:encoding justification}

The encoding procedure mentioned in Subsec. \ref{sec:quantum distillation} can be justified as follows.
The unitary encoding operator can be implemented by applying a certain
quantum circuit, consisting of CNOTs, Hadmard gates, phase gates and SWAP gates. (For example, Wilde gave an encoding algorithm \cite{wilde-2008} to find such a circuit.)

The check matrix of an $[[m,m-r-s]]$ CSS code can be written in the following form (see, e.g., \cite{NC00}):
\[
\sfH=[\sfH_X|\sfH_Z] =   \left(
\begin{array}{ccc|ccc}   \sfI_r& \sfA& \sfB &  0 & 0  &   0\\
                                 0 &  0& 0 &  \sfD & \sfI_s&  \sfF\\
                                 \end{array}\right),
                                 \]
where $\sfI_r$ and $\sfI_s$ are the $r\times r$ and ${s\times s}$ identity matrices, respectively, and {$A$, $B$, $D$, and $F$
are $r\times s$, $r\times (m-r-s)$, $s \times r$, and $s \times (m-r-s)$ binary matrices, respectively.}
($r=s$ in our case.)
Our goal is to apply a sequence of CNOT gates that transform $\sfH$ into
\[
\left(\begin{array}{ccc|ccc}\sfI_r&  0 & 0 & 0 & 0 & 0\\
          0 & 0&  0 &  0 &\sfI_s& 0\\
          \end{array}\right).
\]
Then the reverse of this sequence of CNOTs is our encoding circuit.

This process is like applying Gaussian elimination on $\sfH$.
We first apply a series of CNOTs from the matrices $\sfI_r$ to clear the matrices  $\sfA$ and $\sfB$.
These CNOTs have control qubits on qubit number $1$ to $r$ and target qubits on qubit number $r+1$ to $m$, respectively.
Thus, only the matrix $\sfD$ of $\sfH_Z$ is altered by these CNOTs.
We have
\[
\sfH'= \left(
\begin{array}{ccc|ccc}
\sfI_r& 0& 0&   0&  0&     0\\
        0&  0& 0&   \sfD'&  \sfI_s&  \sfF\\
        \end{array}\right).
\]
Since $\sfH'$ has to satisfy the commutation relations, $\sfD'$ must be $0$.
Thus we have\[
\sfH'= \left(\begin{array}{ccc|ccc}
\sfI_r& 0& 0&   0&  0&     0\\
        0 & 0& 0   & 0 & \sfI_s&  \sfF\\
        \end{array}
        \right).
        \]
Then we apply CNOTs to clear $\sfF$.
These CNOTs have control qubits on qubits number $(r+s+1)$ to $m$ and target qubits on qubits number $(r+1)$ to $(r+s)$, respectively.
$\sfH_X$ is not affected by these CNOTs and we have
\[
\sfH''= \left(\begin{array}{ccc|ccc}\sfI_r&  0 & 0 & 0 & 0 & 0\\
          0 & 0&  0 &  0 &\sfI_s& 0\\
          \end{array}\right)
\]
as required.

\end{document}